\documentclass[aps,prd,preprint,superscriptaddress,tightenlines,nofootinbib]{revtex4}

\usepackage{graphicx}
\usepackage{dcolumn}
\usepackage{bm}
\usepackage{epsfig}
\begin{document}

\preprint{CLNS 03-1832}      
\preprint{CLEO 03-10}         

\title{Cabibbo-Suppressed Decays of $D^+ \to \pi^+ \pi^0, K^+ \bar{K}^0, K^+ \pi^0$}

\author{K.~Arms}
\author{E.~Eckhart}
\author{K.~K.~Gan}
\author{C.~Gwon}
\author{K.~Honscheid}
\author{R.~Kass}
\affiliation{Ohio State University, Columbus, Ohio 43210}
\author{H.~Severini}
\author{P.~Skubic}
\affiliation{University of Oklahoma, Norman, Oklahoma 73019}
\author{S.~A.~Dytman}
\author{J.~A.~Mueller}
\author{S.~Nam}
\author{V.~Savinov}
\affiliation{University of Pittsburgh, Pittsburgh, Pennsylvania 15260}
% ...
\collaboration{CLEO Collaboration} 
\noaffiliation

\date{\today}

\begin{abstract} 
Using a 13.7 fb$^{-1}$ data sample collected with the CLEO II and II.V detectors, 
we report new
branching fraction measurements for two Cabibbo-suppressed decay modes of the $D^+$ meson: 
${\cal B}(D^+\to \pi^+ \pi^0) = (1.31 \pm 0.17 \pm 0.09 \pm 0.09)\times
10^{-3}$ and $ {\cal B}(D^+ \to K^+ \bar{K}^0) = 
(5.24 \pm 0.43 \pm 0.20 \pm 0.34)\times 10^{-3}$
which are significant improvements over past measurements. 
The errors reflect statistical and systematical uncertainties
as well as the uncertainty in the absolute $D^+$ branching fraction scale.  
We also set the first 90\% confidence level upper limit on the
branching fraction of the doubly Cabibbo-suppressed decay mode ${\cal B}(D^+ \to
K^+ \pi^0)  < 4.2\times 10^{-4}$.
\end{abstract}

\maketitle

\begin{figure}
 \begin{center}
  \epsfig{file=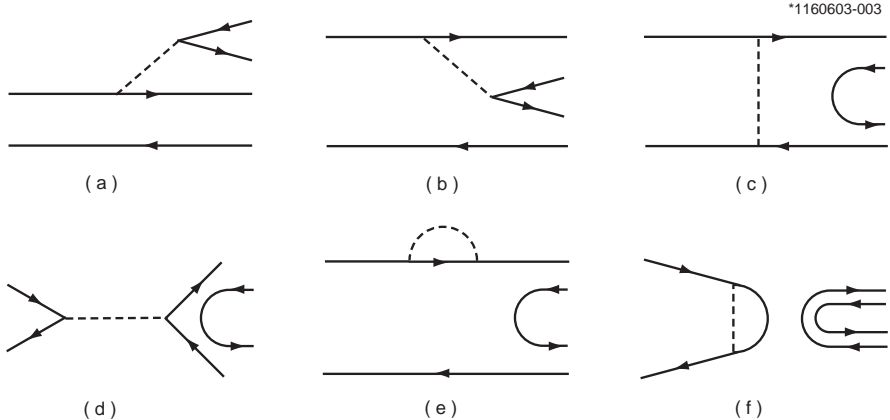,width=4.75in}
  \caption{Six lowest order quark diagrams for a meson 
decaying into two mesons \cite{quarkdiagrams}:
(a) external W-emission, (b) internal W-emission, (c) W-exchange, (d)
W-annihilation, (e) horizontal W-loop, (f) vertical W-loop.  Dashed
lines represent W boson.}
  \label{fig:quarkdiagrams}
 \end{center}
\end{figure}

To lowest order, weak decays of mesons may be described by the six
quark-diagrams shown in Fig. \ref{fig:quarkdiagrams}: external W-emission, internal
W-emission, W-exchange, W-annihilation, horizontal W-loop, and vertical
W-loop~\cite{quarkdiagrams}.  When using these diagrams to
describe processes, dynamical assumptions are often made regarding
the relative size of their amplitudes as well as the nature of
interference terms between diagrams.  Measurements of hadronic decays of $D^+$
mesons give insights into these assumptions
as well as new information on the violation of
$SU(3)$ flavor symmetry ($SU(3)_F$) violation, isospin symmetry, and doubly
Cabibbo-suppressed decays.

$SU(3)_F$ symmetry breaking is of current interest because of
$D^0-\bar{D}^0$ mixing studies; it has been shown that the mass and width differences
($x,y$) of the CP-eigenstates
of neutral $D$ mesons can be generated by the second order $SU(3)_F$
symmetry breaking~\cite{Falk:2001}.  
Understanding the size of these effects may be important to unravel
any non-Standard Model contributions to $D^0\bar{D}^0$ mixing.
Such understanding is only possible if $SU(3)_F$ violating effects
are well-determined. We report new measurements of the decay modes
$D^+ \to \pi^+ \pi^0$ and $D^+
\to K_S^0 K^+$, which are useful for the estimation of
$SU(3)_F$ violating effects in the $D$ meson system.

Predictions based on isospin symmetry are 
generally considered to be more reliable than
$SU(3)_F$ predictions because of the near degeneracy in mass of the {\it u} and 
{\it d} quarks.
Using measurements from
this analysis as well as data from the Particle Date Group (PDG) \cite{pdg}, we determine the isospin amplitudes
and phases for the $D \to \pi \pi$ system.  

Doubly Cabibbo-suppressed decays (DCSD) of charm mesons involve
$c\to d$ and $s\to u$ quark transitions.  Currently, there are only
four measured DCSD decay modes~\cite{pdg}.  Measurements of such
modes will lead to improved understanding of $SU(3)_F$ and other
Standard Model predictions.  Such modes are also important for neutral
$D$-mixing measurements, where a significant background is from DCSD decays.
In this paper we report the first upper
limit on the branching fraction of the DCSD decay $D^+ \to K^+ \pi^0$.  

This analysis uses data collected with two configurations of the CLEO detector at
the Cornell Electron Storage Ring (CESR): CLEO
II~\cite{cleo_kubota} and CLEO II.V~\cite{cleo_hill}. 
The total integrated luminosity of the data sample is 13.7 fb$^{-1}$.
The CLEO detector is a general purpose spectrometer with excellent charged
particle and electromagnetic 
shower energy detection. In CLEO II the momenta of charged particles are
measured with three concentric drift chambers between 5 and 90 cm from the
$e^+e^-$ interaction point. In the CLEO II.V configuration the innermost
drift chamber was replaced by a 3 layer silicon vertex detector. Charged
particles are identified by means of specific ionization measurements
($dE/dx$) in the main drift chamber. The tracking system is surrounded by
a scintillation time-of-flight system and a CsI(Tl) electromagnetic
calorimeter. These detectors are located inside a 1.5 T superconducting
solenoid, surrounded by an iron return yoke instrumented with 
proportional tube chambers for muon identification.

Charged pion and kaon candidates were required to pass minimum
track-quality criteria.  Kaon (pion) candidates had to have a
specific ionization within two (three) standard deviations ($\sigma$) of
that expected for a true kaon (pion). We combined pairs of
electromagnetic showers in the calorimeter to create $\pi^0$
candidates. Candidates with a reconstructed mass 
within 2.5 $\sigma$ of the nominal $\pi^0$ mass were kept for further studies.  
We obtain $K_S^0$
candidates by reconstructing the decay mode $K_S^0 \to \pi^+ \pi^-$.
We required daughter tracks to have an impact parameter in the plane transverse to 
the beam greater
than three times the measurement uncertainty and that the probability of
the $\chi^2$ returned from the vertex fit for pairs of daughter tracks
was required to be greater than 0.001.  $K_S^0$ candidates also had to
have a reconstructed mass within 3.0 $\sigma$ of the nominal
$K_S^0$ mass.

In order to reduce backgrounds, 
we required that $D^{+}$ candidates come from the decay 
$D^{*+}\to D^+\pi^0$, with the mass difference ($\Delta M$) of the
reconstructed $D^{*+}$ and $D^+$ to be within 2.5 $\sigma$ of the known
value \cite{pdg}.  We required all $D^{*+}$ candidates to have a normalized
momentum ($x_{D^*} = {|p_{D^*}|}/{\sqrt{(s/2)^2 - m^2_{D^*}}}$) 
greater than 0.6 and all $D^+$ candidates to
have a $\cos\theta_{h}$ value between $\pm 0.8$, where
$\theta_{h}$ is the angle between the charged daughter track in
the rest frame of the $D^+$ and the reconstructed $D^+$ momentum vector in the rest frame of the
$D^{*+}$ meson.  To insure that we obtained only one $D^+$
candidate per event, we selected candidates with the lowest value for
$$ 
\chi^2 = \frac{(\Delta M - \Delta M_{PDG})^2}{{\sigma_{\Delta M}}^2}+ \sum _i \frac{(m_{\pi^0} -{m_{\gamma\gamma}}^i)^2}
{{\sigma_{\pi^0}}^2},$$
where $i$ indexes the
$\pi^0$ candidates in this decay.  
Given the large uncertainties in absolute $D^+$ branching fractions we present
our results as
ratios of the branching fraction of the decay mode under study
to that of a normalization mode: $D^+ \to K^- \pi^+ \pi^+$ for $D^+
\to \pi^+ \pi^0, K^+ \pi^0$ and $D^+ \to K_S^0 \pi^+$ for $D^+ \to K_S^0
K^+$.  

\begin{figure}
 \begin{center}
  \epsfig{file=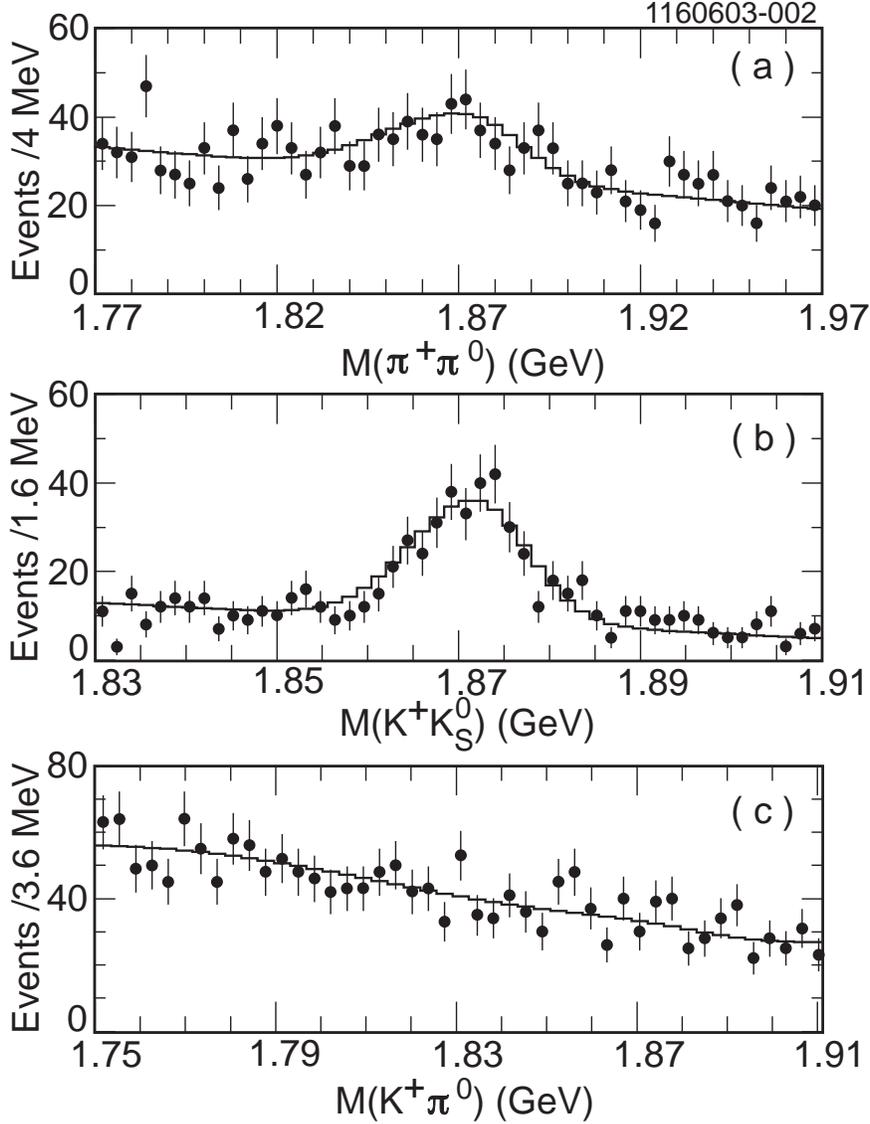,width=4.75in}
  \caption{Invariant mass distributions for (a) $D^+ \to \pi^+ \pi^0$, 
(b) $D^+ \to K^+ K_S^0$, and (c) $D^+ \to K^+ \pi^0$
candidates. The points represent the data and the lines are the projections from the
maximum likelihood fit.}
  \label{fig:data_yields}
 \end{center}
\end{figure}

To extract the yield for each mode, we performed an unbinned maximum
likelihood fit for two components (signal and background) using the
following observables: $m_D$, the mass of the reconstructed $D^+$ meson, 
the normalized momentum of the $D^{*+}$ meson, $x_{D^*}$,
and $\cos \theta_{h}$.
Using a GEANT-based simulation~\cite{geant} of the CLEO detector as well as
sideband data  we determined
probability density functions (PDF) for each observable describing the shape of the
data for signal and background events for each decay mode.  The
probability that a candidate is consistent with signal or background
is given by the product of these PDFs.  The likelihood is given as the
product of these probabilities over all candidates; maximization of
the log of the likelihood gives us the signal and background yields.  Projections of
the likelihood fit to the $D^+$ mass for our three decay modes 
are shown in
Fig.~\ref{fig:data_yields}.  Using simulated signal and background events, 
we measure the efficiency of our
analysis method for each mode, enabling us to determine the total number of
signal events in our data sample for each decay mode.  Table~\ref{tab:fitter_output}
lists raw yields and efficiencies for all decay modes.

\begin{table}[!htb]
 \begin{center} 		
  \caption{Yields from the maximum likelihood fit with statistical 
errors and reconstruction efficiencies.}
  \begin{tabular}{lcc} \hline\hline
      Mode & Yield & Efficiency      \\ \hline
      $\pi^+ \pi^0$     & $  171.3 \pm  22.1$& $(6.20 \pm 0.11)\%$   \\
      $K^+ K_S^0$         & $  277.7 \pm  20.8$& $(4.94 \pm 0.23)\%$   \\
      $K^+ \pi^0$       & $   34.3 \pm  20.9$& $(6.08 \pm 0.22)\%$   \\ \hline
      $K^- \pi^+ \pi^+$ & $12898.0   \pm 156.6$& $(6.74 \pm 0.12)\%$   \\ 
      $\pi^+ K_S^0$       & $ 1434.7  \pm  48.0$& $(4.83 \pm 0.23)\%$   \\ \hline\hline
  \end{tabular}
  \label{tab:fitter_output}
 \end{center}
\end{table}

We considered systematic uncertainties from experimental
resolution, efficiency determination, and PDF parameterization. The first two
contributions are small and the systematic errors are dominated by uncertainties
in the PDF parameterization. We studied this systematic effect for each mode
by simultaneously modifying every PDF parameter within its uncertainty.
We extracted the yield
from the data after each modification to produce a distribution of
yields.  We defined the systematic uncertainty due to PDF
parameterization as the 68\% limits for these distributions.

Combining the systematic error study with the yields and efficiencies given in 
Table~\ref{tab:fitter_output} we obtain the following results:
\begin{eqnarray*}
 \frac{{\cal B}(D^+ \to \pi^+ \pi^0)}{{\cal B}(D^+ \to K^- \pi^+
\pi^+)} & = & 0.0144 \pm 0.0019 \pm 0.0010 \\
 \frac{{\cal B}(D^+ \to K^+ K_S^0)}{{\cal B}(D^+ \to \pi^+ K_S^0)} & = & 0.1892 \pm 0.0155 \pm 0.0073 \\
 \frac{{\cal B}(D^+ \to K^+ \pi^0)}{{\cal B}(D^+ \to K^- \pi^+ \pi^+)}
& = & 0.0029 \pm 0.0018 \pm 0.0009
\end{eqnarray*}
\noindent where the first error is statistical and the second error is
systematic.  The results supersede previous CLEO measurements 
\cite{Geiser:1993pt,Kim:1997km}.

In order to determine the absolute branching fractions, we combine
our results with the PDG values~\cite{pdg} of 
${\cal B}(D^+ \to K^- \pi^+ \pi^+) = (9.1 \pm 0.6)\%$ and ${\cal
B}(D^+ \to \pi^+ \bar{K}^0) = (2.77 \pm 0.18)\%$
and find
\begin{eqnarray*}
 {\cal B}(D^+ \to \pi^+ \pi^0) & = & (1.31 \pm 0.17 \pm 0.09 \pm 0.09)\times 10^{-3} \\
 {\cal B}(D^+ \to K^+ \bar{K}^0) & = & (5.24 \pm 0.43 \pm 0.20 \pm 0.34)\times 10^{-3} \\
 {\cal B}(D^+ \to K^+ \pi^0) & = & (2.64 \pm 1.64 \pm 0.82 \pm 0.17)\times 10^{-4} 
\end{eqnarray*}
\noindent where the third listed uncertainty comes from the error of
branching fractions of the normalization modes.

With no significant signal being observed 
for the doubly Cabibbo-suppressed decay $D^+ \to K^+\pi^0$ we
determined the 90\% confidence level upper limit for this branching fraction.
Our method for obtaining the upper limit involved
creating 1000 new data sets with the same number of signal and background
events as our data sample.
In order to include systematic
uncertainties in our upper limit, we also modified the PDF parameters
in the manner described for our branching fraction calculation.  Using
this method, our upper limit is  
$${\cal B}(D^+\to K^+ \pi^0) < 4.2
\times 10^{-4} \; {\rm at\; 90\% \;C.L.}$$

In the limit of $SU(3)_F$, the following ratio is expected to be
unity~\cite{chau}
\[R_1 = 2\times \left | \frac{V_{cs}}{V_{cd}} \right |^2
\frac{\Gamma(D^+ \to \pi^+ \pi^0)}{\Gamma(D^+ \to \bar{K}^0 \pi^+)}\]
\noindent 
where the $V_{cs}$ and $V_{cd}$ arise because of the different quark transitions
in the two decays and the factor of 2 arises because of the $\sqrt{{1}/{2}}$ term 
in the normalization of the $\pi^0$ wavefunction.
Using $|V_{cs}|/|V_{cd}| = 4.45 \pm 0.32$ \cite{pdg},
the yields and efficiencies (Table \ref{tab:fitter_output})
obtained from our analysis and combining statistical and systematical uncertainties
in quadrature, we find 
$$
R_1 = 1.84 \pm 0.38
$$ 
slightly inconsistent with theoretical
expectations that $SU(3)_F$ symmetry breaking effects are about 30\%.

It is believed that in the $D$ meson system the interference between  external
and internal W-emission decay amplitudes (Figure \ref{fig:quarkdiagrams})
is destructive. In order to test this
assumption we calculate the ratio
\[R_2 = \frac{1}{2}\times\frac{\Gamma(D^+ \to  K^+\bar{K}^0)}{\Gamma(D^+ \to \pi^+ \pi^0)}
= \frac{\Gamma(D^+ \to K^+ K_S^0 )}{\Gamma(D^+ \to \pi^+ \pi^0)}. \]
\noindent 
which in case of destructive interference should be greater than 1.
Besides a small contribution from the W-annihilation diagram~\cite{chau}
the decay in the numerator, $D^+ \to K^+ K_S^0$, can be described using an
external W-emission diagram. Whereas both the external and the internal
W-emission amplitudes contribute to the decay in the 
denominator, $D^+ \to \pi^+\pi^0$.
Experimentally, we find using our yields and efficiencies from
Table \ref{tab:fitter_output}
$$
R_2 = 2.03 \pm 0.32
$$
indicating that the interference
between external and internal W-emission is indeed destructive.

Final state interactions (FSI) are significant in charm decays. Using our measurement for $D^+ \to \pi^+ \pi^0$ and the PDG
values for $D^0 \to \pi^+ \pi^-, \pi^0 \pi^0$~\cite{pdg} we can gain some insights on
these effects by determining 
isospin amplitudes and phases for the $D \to \pi \pi$ system.  
The $\pi \pi$ final state 
may have an isospin value of 0 or 2.  Writing the amplitudes for
the I = 0 state as $A_0$ and the I = 2 state as $A_2$, we obtain the
following relation:
\[ \left |\frac{A_2}{A_0}\right |^2 =
\frac{\Gamma^{+0}}{\frac{3}{2}(\Gamma^{+-} + \Gamma^{00}) -
\Gamma^{+0}} \]
\noindent where $\Gamma^{ab} = \Gamma(D^+ \to \pi^a \pi^b)$ and $a,b$
represent the charges of the pions.  Since isospin amplitudes are
complex, measuring the phase between them is necessary to obtain
full information about the amplitudes.  The phase is written as
\[ \cos\delta =
\frac{3\Gamma^{+-} - 6\Gamma^{00} + 2\Gamma^{+0}}{4\sqrt{2\Gamma^{+0}}\sqrt{\frac{3}{2}(\Gamma^{+-} + \Gamma^{00}) -\Gamma^{+0}}}. \]
\noindent We find $|A_2/A_0| =
0.421\pm 0.044$ and $\cos\delta = 0.042 \pm 0.195$. These results supersede a 
previous CLEO measurement \cite{Geiser:1993pt}. The large relative phase between
the isospin amplitudes indicate that there are significant FSI effects in the $D \to \pi
\pi$ system confirming our earlier results \cite{Geiser:1993pt}. A similar observation has 
been made recently by the FOCUS collaboration \cite{focus}.

In summary, we have obtained measurements for two singly Cabibbo-suppressed $D^+$ decay
modes: ${\cal
B}(D^+\to \pi^+ \pi^0) = (1.31 \pm 0.17 \pm 0.09 \pm 0.09)\times
10^{-3}$ and $ {\cal B}(D^+ \to K^+ \bar{K}^0) = 
(5.24 \pm 0.43 \pm 0.20 \pm 0.34)\times 10^{-3}$.  
We also present an upper limit on the DCSD mode
${\cal B}(D^+\to K^+ \pi^0) < 4.2 \times 10^{-4}$ at the 90\% C.L.
Our experimental measurements confirm the destructive nature of the
interference term between the external and internal W-emission
diagrams and indicate significant $SU(3)_F$ symmetry breaking.
An isospin analysis shows that FSI effects are important for hadronic decays of
$D$ mesons.
%

We gratefully acknowledge the effort of the CESR staff 
in providing us with
excellent luminosity and running conditions.
This work was supported by 
the National Science Foundation,
the U.S. Department of Energy,
the Research Corporation,
and the 
Texas Advanced Research Program.

\end{document}